\documentclass[conference]{IEEEtran}

\usepackage{cite}
\usepackage{amssymb,amsfonts,amsmath,amsthm,amscd,dsfont,mathrsfs}
\usepackage{graphicx,float,psfrag,epsfig,color,url}
\usepackage{latexsym}
\renewcommand{\theenumi}{\roman{enumi}.}

\newtheorem{theorem}{Theorem}[section]
\newtheorem{lemma}[theorem]{Lemma}
\newtheorem{corollary}[theorem]{Corollary}
\newtheorem{definition}{Definition}[section]
\newtheorem{remark}[definition]{Remark}

\newcommand{\comment}[1]{}

\def\prob{{\mathbb P}}

\newcommand{\reals}{{\mathds R }}

\def\Var{{\rm Var}}
\def\cA{{\cal A}}
\def\E{{\mathbb E}}

\def\prob{{\mathbb P}}

\def\de{{\rm d}}

\def\sign{{\rm sign}}

\def\hM{\widehat{M}}
\def\tA{{\tilde{A}}}
\def\tr{{\rm Tr}}
\def\supp{{\rm supp}}

\begin{document}

\title{Information Theoretic Limits on\\ Learning Stochastic Differential Equations}

\author{\IEEEauthorblockN{Jos\'e Bento and Morteza Ibrahimi}
\IEEEauthorblockA{Department of Electrical Engineering\\
Stanford University}
\and
\IEEEauthorblockN{Andrea Montanari}
\IEEEauthorblockA{Department of Electrical Engineering and\\
Department of Statistics\\
Stanford University}}

\maketitle

\begin{abstract}
Consider the problem of learning the drift coefficient of a
stochastic  differential equation from a sample path.
In this paper, we assume that the drift is parametrized by a 
high-dimensional vector. We address the question of how 
long the system needs to be observed in order to learn this vector
of parameters.
We prove a general lower bound on this time complexity
by using  a characterization of mutual information as time integral
of conditional variance, due to Kadota, Zakai, and Ziv. 
This general lower bound is applied to specific classes of linear and
non-linear stochastic differential equations.
In the linear case, the problem under consideration is the one 
of learning a matrix of interaction coefficients. 
We evaluate our lower bound for ensembles of sparse and dense
random matrices. The resulting estimates match the
qualitative behavior of upper bounds achieved by computationally 
efficient procedures.
\end{abstract}
%
%
\section{Introduction}

Consider a continuous-time stochastic process $\{x_t\}_{t \geq 0}$, 
that is defined by a stochastic differential equation (SDE) of the form
\begin{eqnarray}
\de x_t = 
F(x_t;A) \, \de t + \de b_t \, , \label{eq:BasicModel}
\end{eqnarray}
where $x_t \in\reals^p$, $b_t$ is a $p$-dimensional standard Brownian
motion and the \textit{drift coefficient} $F(x_t;A) =
[F_1(x_t;A),...,F_p(x_t;A)] \in \reals^p$, is a function of $x_t$
parametrized by $A$, which is an unknown high-dimensional vector.

In this paper we consider the problem of learning information about
the vector of parameters $A$ 
from the observation of a sample trajectory $X^T \equiv \{x_t\}^T_{t =
  0}$.  More precisely, we consider the high dimensional case (where the dimensions of $A$ and $x_t$ are large) and investigate what is
the minimum  time length $T$ we need to observe the system 
in order to be able to recover $A$, with some confidence. 

Models based on SDE's play a crucial role in several domains of
science and technology, ranging from chemistry to finance.
As an example, gene regulatory networks can be  modeled
by systems of non-linear stochastic differential equations,
whose variables encode concentrations of certain gene expression 
products (e.g. proteins) \cite{BioBook}. Complex chemical networks 
are also described by SDE's that can involve
hundreds of reactants \cite{Gillespie,Higham}. 
The problem of learning the parameters (reaction coefficients) of 
such an SDE or simply reconstructing the underlying network structure
(i.e. which parameters are non-vanishing)
plays crucial role in this context \cite{ToniEtAl}.

An important subclass of  models consists in linear SDE's, 
whereby the drift is a linear function of $x_t$, namely 
$F(x_t;A) = Ax_t$ with $A \in \reals^{p\times p}$.
This can be a good approximation for many systems near a stable 
equilibrium.
Linear SDE's are a special case of a broader class 
for which the drift is a linear combination of a finite set of 
basis functions $F(x_t) = [f_1(x_t), f_1(x_t), \dots, f_m(x_t)]$, with 
$f_i:\reals^p \rightarrow \reals$. The drift is then given as $ F(x_t;A) = AF(x_t)$,
with $A \in \reals^{p \times m}$.
As an example, within models of chemical reactions, the drift is a
low-degree polynomial. For instance, the reaction ${\sf A}+2{\sf B}\to{\sf C}$
is modeled  as $\de x_{{\sf C}} = k_{{\sf C,AB}}x_{{\sf A}}x_{{\sf
  B}}^2\de t+ \de b_{{\sf C}}$ where $x_A$, $x_B$ and $x_C$ denote the concentration of the species $A$, $B$ and $C$ respectively, and $\de b_C$ is a noise term affecting the measurement of $x_C$. In order to learn a model of this type, one
can consider a  basis of functions that contain all monomials 
up to a maximum degree.
%
%
\subsection{Illustration}

As an illustration, consider a system of $p$ masses  in
$\reals^d$ connected by springs. 
Let $C^0$ be the corresponding adjacency matrix,
i.e. $C^0_{ij}=1$ if and only if masses $i$  and $j$ are connected, and
$D^0_{ij}$  be the rest length of the spring $(i,j)$. Assuming unit
masses and unit elastic coefficients, the dynamics of this system in
the presence of external noisy forces 
can be modeled by the following damped Newton equations
\begin{align} 
&\de v_t = -\gamma v_t \de t-\nabla U(q_t)\, \de t  +\sigma\, \de b_t, \label{eq:general_ms_dynamics}\\
&\de q_t = v_t \de t\, ,\label{eq:general_ms_dynamics_2}\\
& U(q) \equiv
\frac{1}{2}\sum_{(i,j)}C^0_{ij}(\|q^{(i)}-q^{(j)}\|-D^0_{ij})^2\, ,\nonumber
\end{align}
where $q_t =
(q^{(1)}_t,\dots,q^{(p)}_t)$, $v_t = (v^{(1)}_t,\dots,v^{(p)}_t)$, and
$q^{(i)}_t, v^{(i)}_t\in \reals^d$ denote the position and velocity of mass $i$ at time $t$. 
This system of SDE's can be written in the form \eqref{eq:BasicModel}
by letting $x_t = [q_t, v_t]$ and $A =[C^0, D^0]$ . A
straightforward calculation shows that 
the drift $F(x_t;A)$ can be further written as a linear combination
of the following basis of non-linear functions
\begin{align}
F(x_t) = \Big[\{v^{(i)}_t\}_{i \in [p]}, \{ \Delta^{(ij)}_t\}_{i,j \in [p]},
\Big\{ \frac{\Delta^{(ij)}_t}{\|\Delta^{(ij)}_t\|} \Big\}_{i,j \in [p]} \Big],
\end{align}
where $\Delta^{(ij)}_t = q^{(i)}_t-q^{(j)}_t$ and $[p] =
\{1,\dots,p\}$. In many situations only specific properties of the
parameters are of interest, for instance one might be interested
only in the network structure in the present example. 

\begin{figure}
\begin{center}
\includegraphics[scale = 0.27]{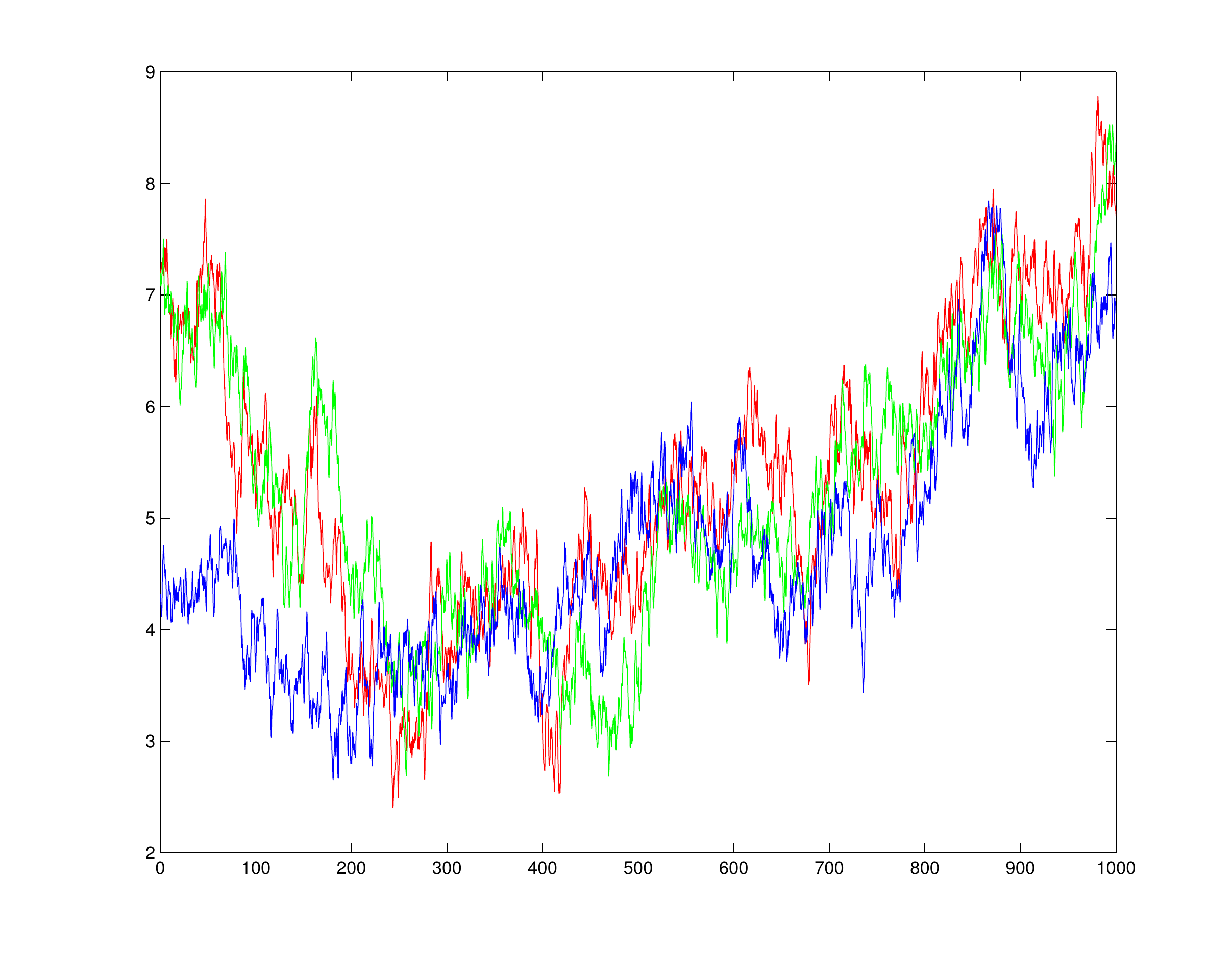}
\put(-45,21){Time}
\put(-160,122){Displacement}
\caption{Evolution of the horizontal component of the position of
  three masses in a system with $p=36$ masses interacting via elastic
  springs (cf. Fig.~\ref{fig:ms_network_reconstrunct} for the network
  structure).  The time interval is here $T=1000$. All the springs have
  rest length $D_{ij}=1$, the damping coefficient is $\gamma=2$,
 cf. Eq.~(\ref{eq:general_ms_dynamics}), 
and the noise variance is $\sigma^2 =0.25$.}\label{fig:orig_ms_network}
\end{center}
\end{figure}

\begin{figure}
\begin{center}
\includegraphics[scale = 0.20]{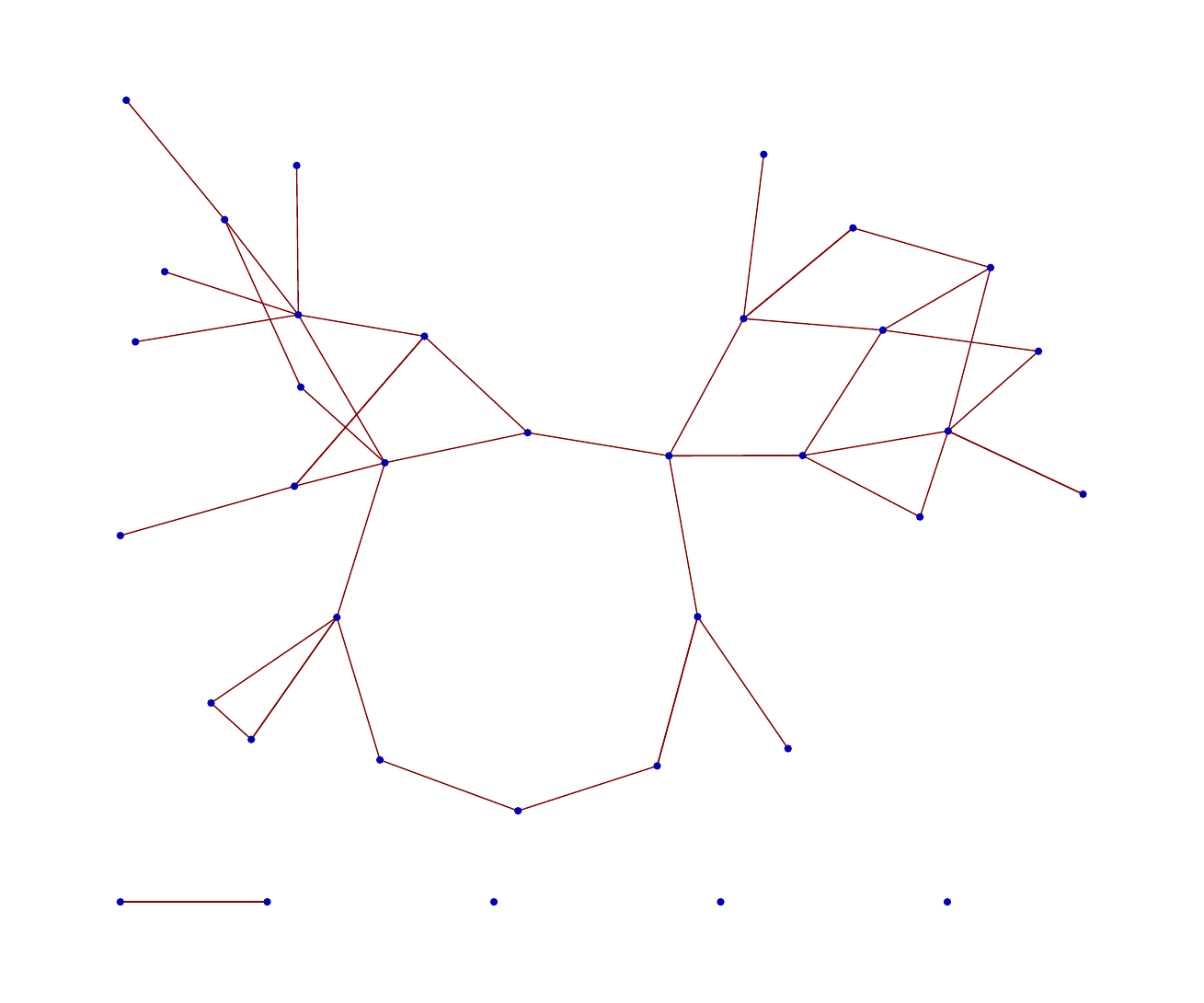}
\includegraphics[scale = 0.20]{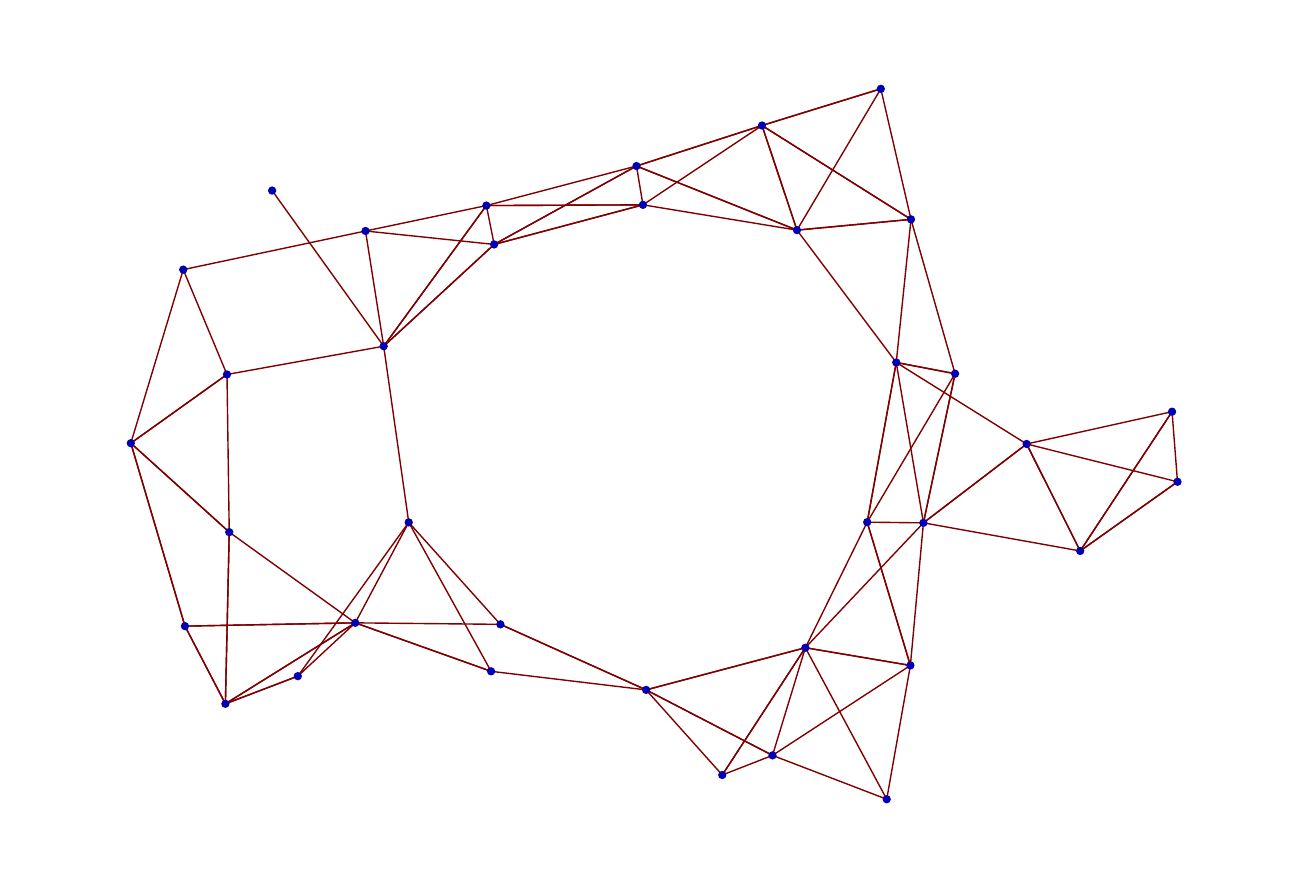}
\includegraphics[scale = 0.20]{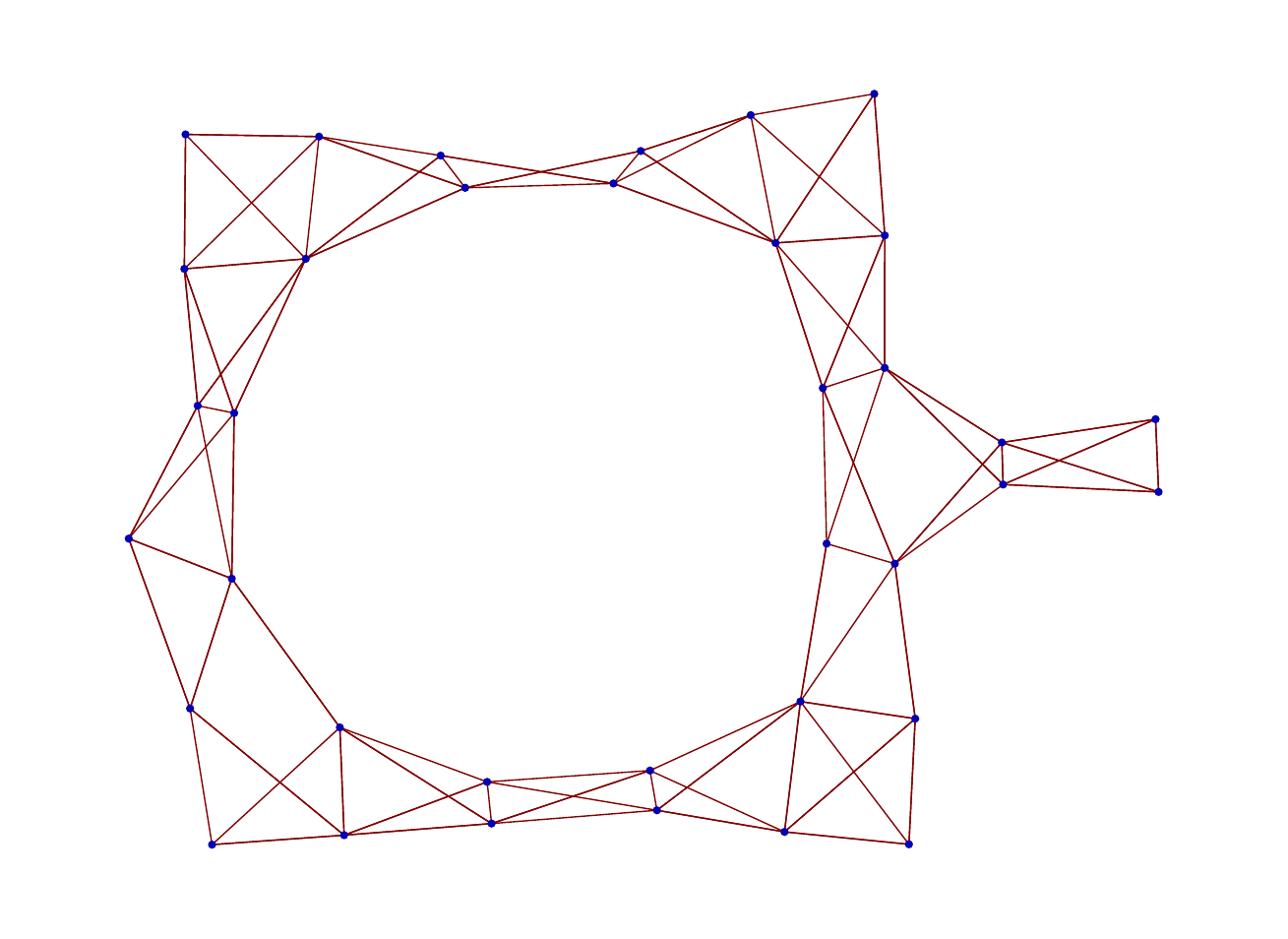}
\includegraphics[scale = 0.20]{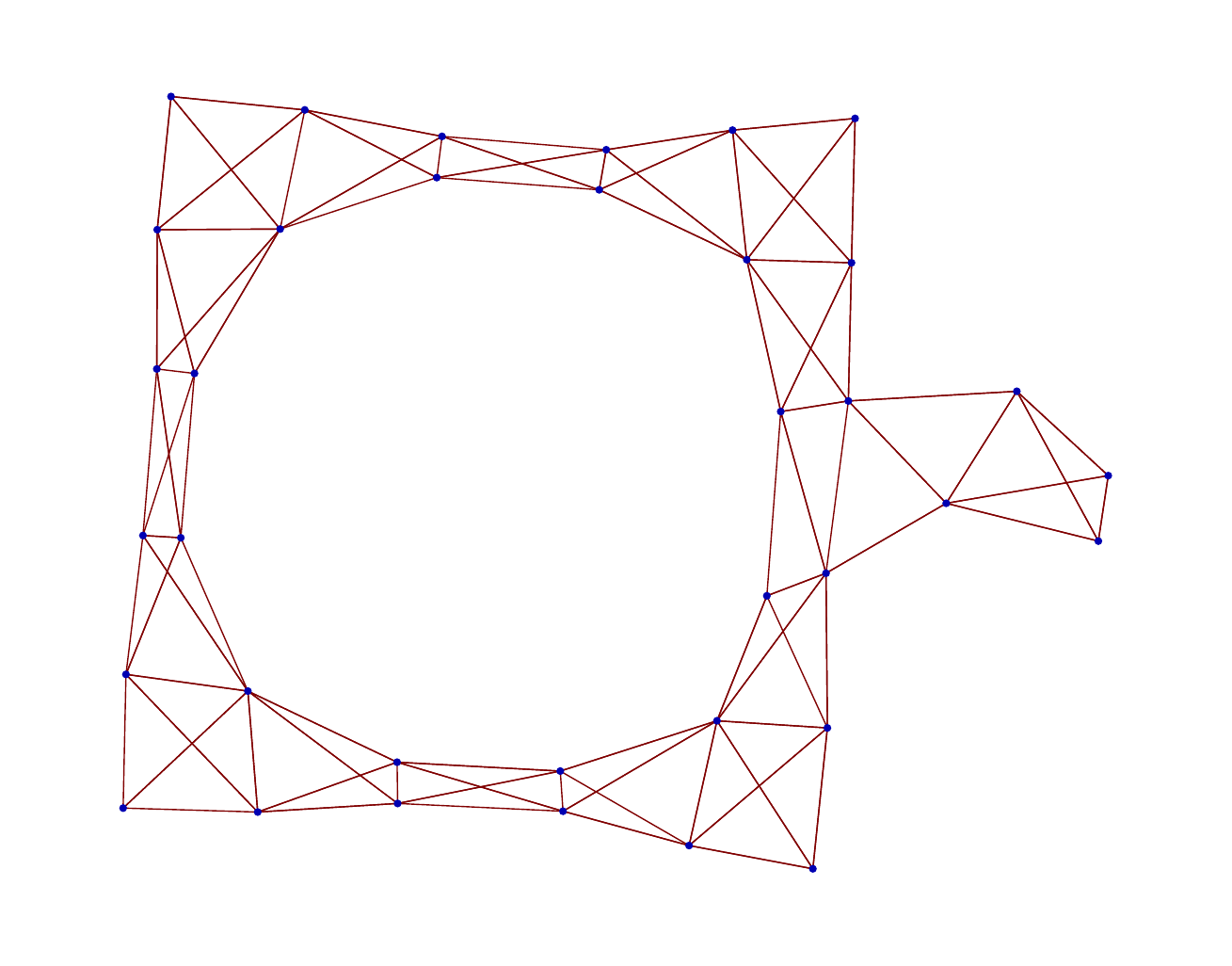}
\includegraphics[scale = 0.20]{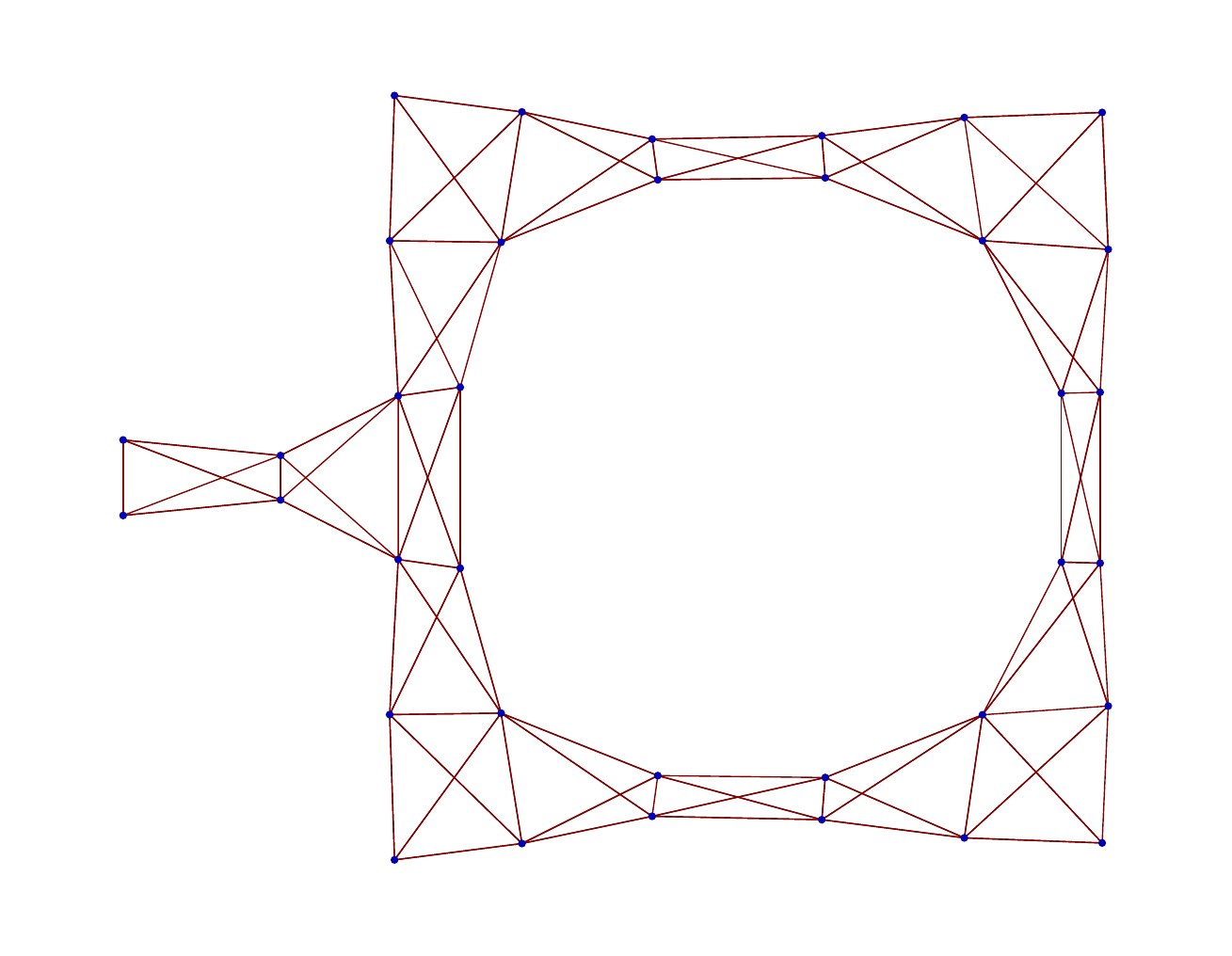}
\caption{From left to right and top to bottom: structures  reconstructed using the algorithm of
  \cite{bento2010learning} with 
  observation time $T=500$, $1500$, $2500$, 
$3500$ and $4500$. For $T=4500$ exact reconstruction is achieved.}\label{fig:ms_network_reconstrunct}
\end{center}
\end{figure}

Figure \ref{fig:orig_ms_network} shows the trajectories of three
masses in a two-dimensional network of 36 masses and 90 springs evolving according
to Eq.~(\ref{eq:general_ms_dynamics}) and Eq.~(\ref{eq:general_ms_dynamics_2}).
How long does one need to observe these (and the other masses) trajectories 
in order to learn the structure of the underlying network? Figure 
\ref{fig:ms_network_reconstrunct}  reproduces the network structure 
reconstructed using the algorithm of \cite{bento2010learning}
for increasing observation intervals $T$. The inferred structure
converges to the actual one only if $T$ is large enough.

\subsection{Related Work}

Over the last few years, a significant effort has been devoted to
developing methods  and sample complexity 
bounds for learning graphical models from data. 
Particular effort was devoted to learning sparse graphical models
using convex regularizations that promote sparsity. 
Well known examples in the context of Gaussian graphical models
include the \emph{graphical} LASSO \cite{friedman2008sparse}
and the pseudo-likelihood method of \cite{Buhlmann}.
These papers assume that the data are i.i.d. samples from a
high-dimensional Gaussian distribution. However in many cases 
samples are produced by an underlying dynamical process and the
i.i.d. assumption is unrealistic. 

In \cite{bento2010learning}, a convex regularization method was
developed to learn linear SDE's with a sparse network structure
from data. The upper bounds on the sample
complexity proved in \cite{bento2010learning} match in several cases
the lower bounds developed here.
The related topic of learning graphical models for
autoregressive processes was studied recently in 
\cite{Songsiri1,Songsiri2}. These papers propose a convex relaxation
different from the one of \cite{bento2010learning}, without however
developing estimates on the sample complexity for  
model selection.

Finally, a substantial literature addresses various questions
related to learning SDE's \cite{Basawa,Stuart,Higham}. 
However this line of work did not yield quantitative estimates on
the scaling of sample complexity
with the problem dimensionality.
%
%
\section{Main Results}

Without loss of generality, assume that the parameter $A$ is a random variable
chosen with some unknown prior distribution $\prob_{A}$ (subscript
will be often omitted). We are interested in a specific property 
of $A$ that is given by a function $A\mapsto M(A)$.
Unless specified otherwise $\prob$ and $\E$ denote probability
and expectation with respect to the joint law of $\{x_t\}_{t\ge 0}$
and $A$. As mentioned above $X^T\equiv \{x_t\}_{0\le t\le T}$
will denote the trajectory up to time $T$.
Also, we define the variance of a vector-valued random
variable as the sum of the variances over all components, i.e.,
\begin{equation}
\Var_{A|X^t } (F(x_t;A)) = \sum^p_{i = 1}  \Var_{A|X^t } (F_i(x_t;A)).
\end{equation}

Our main tool is the following general lower bound,
that follows from an identity between mutual information and the integral
of conditional variance proved by Kadota, Zakai and Ziv \cite{kadotamutinfo}.
\begin{theorem}\label{th:main_lbound}
Let $\hM_T(X^T)$ be an estimator of $M(A)$ based on
$X^T$. If  
$\prob(\hM_T(X^T) \neq M(A) ) <\frac{1}{2}$ then
\begin{equation} \label{eq:main_bound}
T \geq \frac{H(M(A)) - 2 I(A;x_0)}{\frac{1}{T} \int^T_0
  \E_{X^t}  \{ \Var_{A|X^t } (F(x_t;A)) \} \de t}\, .
\end{equation}
\end{theorem}
\begin{IEEEproof}
Equation \eqref{eq:BasicModel} can be regarded as describing a white Gaussian channel with feedback where $A$ denotes the message to be transmitted. For this scenario, Kadota et al. \cite{kadotamutinfo} give the following identity for the mutual information between $X^T$ and $A$ when the initial condition is $x_0 = 0$,
\begin{equation}
I(X^T;A) = \frac{1}{2} \int^T_0 \E_{X^t}  \{ \Var_{A|X^t } (F(x_t;A)) \} \de t.
\end{equation}
For the general case where $x_0 \neq 0$ and might depend on $A$ (if for example $x_0$ is the stationary state of the system) we can write $I(X^T;A) = I(x_0;A) + I(X^T;A | x_0)$ and apply the previous identity to $I(X^T;A | x_0)$. Taking into account that $I(\hM_T(X^T));M(A)) \leq I(X^T;A)$ and making use of Fano's inequality $I(\hM_T(X^T));M(A)) \geq \prob(\hM_T(X^T) = M(A) ) H(\hM_T(X^T)))$ the results follows.
\end{IEEEproof}

The bound in Theorem \ref{th:main_lbound} is often too complex to be
evaluated.  Instead, the following corollary provides a more easily
computable
 bound.
\begin{corollary}\label{th:main_simpler_lbound}
Assume that the process $\{x_t\}_{t\ge 0}$ is stationary.
Let $\hM_T(X^T)$ be an estimator of $M(A)$ based on $X^T$. If  $\prob(\hM_T(X^T) \neq M(A) ) <\frac{1}{2}$ then
\begin{equation} \label{eq:main_simpler_bound}
T \geq \frac{H(M(A)) - 2 I(A;x_0)}{     \E_{x_0}  \{ \Var_{A|x_0 } (F(x_0;A)) \}} .
\end{equation}
\end{corollary}
\begin{IEEEproof}
Since conditioning reduces variance, we have $\E_{X^t}  \{
\Var_{A|X^t } (F(x_t;A)) \} \leq \E_{x_t}  \{ \Var_{A|x_t }
(F(x_t;A)) \} $. Using stationarity, we have 
$\E_{x_t}  \{ \Var_{A|x_t }
(F(x_t;A)) \}=\E_{x_0}  \{ \Var_{A|x_0 }
(F(x_t;A)) \} $, 
 which simplifies \eqref{eq:main_bound} to \eqref{eq:main_simpler_bound}.
\end{IEEEproof}
In the rest of this section, we apply this lower bound to special
classes of SDE's. In all of our applications it is understood that the 
process $\{x_{t}\}_{t\ge 0}$ is stationary.
%
%
\subsection{Learning Sparse Linear SDE's}
Consider the linear SDE, 
\begin{equation}\label{eq:linear_model}
\de x_t = A x_t \de t + \de b_t.
\end{equation}
The goal is to learn the interaction matrix $A \in \reals^{p\times p}$. 
The first two theorems stated below provide lower bounds for sample complexity $T$, for the two regimes of sparse and dense matrices. 
Throughout this paper $Q^*$ will denote the transpose of matrix $Q$.  
Given a matrix $Q$, its $\supp(Q)$ is the $0-1$ matrix such that 
$\supp(Q)_{ij}=1$ if and only if $Q_{ij}\neq 0$. 
Its `signed support' $\sign(Q)$ is  the matrix such that 
$\sign(Q)_{ij}=\sign(Q_{ij})$ if $Q_{ij}\neq 0$ and 
$\sign(Q)_{ij}=0$ otherwise.

Define the class of matrices $\cA^{(S)} \subset \reals^{p\times p}$ by 
letting $A \in \cA^{(S)}$ if and only if
\begin{itemize}
\item[(i)]  $A$ has at most $k$ non-zero elements per row, $k \geq 3$,
\item[(ii)] $\min_{ij} |A_{ij}| > a_{\min}$,
\item[(ii)] Letting $\lambda_{\min}(Q)$ denote the smallest eigenvalue of 
matrix $Q$, $\lambda_{\min}(-(A+A^*)/2) \ge \rho > 0$.
\end{itemize}
The next theorem provides a lower bound on the time complexity of
learning the signed support of models from the class $\cA^{(S)}$.
\begin{theorem} \label{th:linear_lbound_sparse}
Let $M(A) = \sign(A)$ be the signed support of $A$ and $\hM_T(X^T)$ an estimator of $M(A)$ based on $X^T$. There is a constant $C(k)$ such that,
for all $p$ large enough,
if $\sup_{A \in \cA^{(S)}} \prob_{X^T|A} (M(A) \neq \hM_T(X^T)) <\frac{1}{2}$ then 
\begin{align}
T >  \frac{C(k)}{a_{\min}} \max \{ \rho/a_{\min},1\} \log (p).
\end{align}
\end{theorem}

\subsection{Learning Dense  Linear SDE's}

A different regime of interest in learning the network of interactions for a linear SDE's is the case of dense matrices. As we shall see shortly, this regime exhibits fundamentally different behavior in terms of sample complexity compared to the regime of sparse matrices. 

Let $\cA^{(D)} \subset \reals^{p\times p}$ be the set of matrices with the following properties: $A \in \cA^{(D)} $ if and only if, 
\begin{itemize}
\item[(i)] $ a_{\min}\leq |A_{ij}| p^{1/2} \leq a_{\max}$.
\item[(ii)] $\lambda_{\min}(-(A+A^*)/2) \ge \rho > 0$.
\end{itemize}
The second theorem provides a lower bound for learning the signed support of models from class $\cA^{(D)}$.
\begin{theorem}\label{th:linear_lbound_dense}
Let $M(A) = \sign(A)$  be the signed support of $A$ and $\hM_T(X^T)$ an estimator of $M(A)$ based on $X^T$.  There exists a constant $C$ such that, for all $p$ large enough, if $\sup_{A \in \cA^{(D)}} \prob_{X^T|A} (M(A) \neq \hM_T(X^T)) <\frac{1}{2}$ then 
\begin{align}
T >  \frac{C}{a_{\min}}\max \{\rho / a_{\min}, 1\} p.\label{eq:lbound_dense}
\end{align}
\end{theorem}

Together with the upper bounds from  \cite{bento2010learning},
Theorem \ref{th:linear_lbound_sparse} establishes that the time
complexity of learning sparse linear SDE's  is $T=\Theta(\log(p))$.
Further, this task can be performed efficiently using
$\ell_1$ penalized least squares  \cite{bento2010learning}.
On the other hand, Theorem \ref{th:linear_lbound_dense} 
implies a dramatic dichotomy. The time complexity of learning 
dense linear SDE's is at least linear in $p$
(and indeed matching upper bounds can be proved in this
case as well \cite{underpreparation}).

%
%
\subsection{Learning Non-Linear SDE's}

In this section we assume that the observed samples $X^T$ come from a stochastic process driven by a general SDE of the form \eqref{eq:BasicModel}.

In what follows, $v^{(i)}$ denotes the $i^{th}$ component of vector $v$. For example, $x^{(3)}_2$ is the $3^{th}$ component of the vector $x_t$ at time $t = 2$. $JF(\,\cdot\, ;A)\in\reals^{p\times p}$ will denote the Jacobian of the 
function $F(\,\cdot\, ;A)$.

For fixed $L$, $B$ and $D\ge 0$, 
define the class of functions 
$\cA^{(N)}=\cA^{(N)}(L,B,D)$ by letting $F(x;A) \in \cA^{(N)}$ if and only if
\begin{itemize}
\item[(i)] the support of $JF(x;A)$ has at most $k$ non-zero entries for every $x$,
\item[(ii)] the covariance matrix for the stationary process, $\Sigma_{\infty}$, satisfies $\lambda_{\min}(\Sigma_{\infty}) \geq L$,
\item[(iii)] $\Var_{x_0|A}(x^{(i)}_0) \leq B \; \forall i$,
\item[(iv)] $|\partial F_i(x;A) / \partial x^{(j)} |\leq D$ for all
$x\in\reals^p$ $i,j\in [p]$.
\end{itemize}

For simplicity we write $F(x;A) \in \cA^{(N)}$ by $A \in \cA^{(N)}$.

\begin{theorem} \label{th:non_linear_lbound}
Let $M(A)$ be the smallest support for which $\text{supp}(JF(x;A)) \subseteq M(A) \; \forall x$. If $\hM_T(X^T)$ is an estimator of $M(A)$ based on $X^T$ and $\sup_{A \in  \cA^{(N)}} \prob_{X^T | A} (\hM_T(X^T) \neq M(A)) < 1/2$ then
\begin{equation}
T > \frac{k \log p/k   - \log B/L }{C + 2k^2 D^2 B}.
\end{equation}
In the above expression $C = \max_{i \in [p]}\E \{F_i(\E_{x_0|A}(x_0);A)\}$.
\end{theorem}

\begin{remark}
Note that the assumption that $F$ is Lipschitz is not very strong in the sense that it is usually required for existence and uniqueness of a solution of the SDE \eqref{eq:BasicModel} with finite expected energy, \cite{oksendal2003stochastic}.
\end{remark}
%
%
\section{Proofs and technical lemmas}

In this section we prove Theorems \ref{th:linear_lbound_sparse}
to \ref{th:non_linear_lbound}. Throughout, $\{x_t\}_{t\ge 0}$ 
is assumed to be a stationary process. It is immediate to check
that under the assumptions of the Theorems \ref{th:linear_lbound_sparse} and \ref{th:linear_lbound_dense}, the SDE admit a unique stationary measure, with bounded covariance.
We let  $\Sigma_{\infty} = \E\{x_0x_0^*\} - \E\{x_0\}(\E\{x_0\})^* =\E\{x_tx_t^*\} - \E\{x_t\}(\E\{x_t\})^*$ denote this
covariance. 

\subsection{A general bound for linear SDE's}

 Before passing to the actual proofs, it 
is useful to establish a general bound for linear SDE's  
(\ref{eq:linear_model}) with symmetric
interaction matrix $A$. 
\begin{lemma} \label{th:mut_inf_bound_linear}
Assume that  $\{x_t\}_{t\ge 0}$ is a stationary process
generated by the linear SDE (\ref{eq:linear_model}), with $A$ symmetric.
Let $\hM_T(X^T)$ be an estimator of $M(A)$ based on $X^T$. 
If  $\prob(\hM_T(X^T) \neq M(A) ) <\frac{1}{2}$ then
\begin{equation} \label{eq:bound_linear}
T \geq \frac{H(M(A)) - 2 I(A;x_0)}{ \frac{1}{2}\tr
\{ \E \{-A \}-   (\E\ \{ -A^{-1} \})^{-1} \} \}} .
\end{equation}
\end{lemma}
\begin{IEEEproof}
The bound follows from Corollary \ref{th:main_simpler_lbound}
after showing that    $\E_{x_0}  \{ \Var_{A|x_0 } (Ax_0))\le (1/2)\tr
\{ \E \{-A \}-   (\E\ \{ -A^{-1} \})^{-1} \}$.
First note that
\begin{align}
  \E_{x_0}  \{ \Var_{A|x_0 } (Ax_0) \}  = \E_{x_0} ||  A x_0  - \E_{A|x_0 } (Ax_0|x_0) ||^2_2.\label{eq:norm2_err}
\end{align}
The quantity in \eqref{eq:norm2_err} can be thought of as the $\ell_2$-norm error of estimating $A x_0$ based on $x_0$, using $\E_{A|x_0 }(Ax_0|x_0)$. Since conditional expectation is the minimal mean square error estimator, replacing 
$\E_{A|x_0 } (Ax_0|x_0)$ by any estimator of $Ax_0$ based on $x_0$ gives an upper bound for the expression in \eqref{eq:norm2_err}. We choose as an estimator a linear estimator , i.e., an estimator in the form $B x_0$ where $B = (\E_A A \Sigma_{\infty}) (\E_A \Sigma_{\infty})^{-1}$,
\begin{align} \label{eq:err_linear_est}
 &\E_{x_0} ||  A x_0  - \E_{A|x_0 } (Ax_0|x_0) ||^2_2 \leq  \E_{x_0} ||  A x_0  - B x_0 ||^2_2  \nonumber \\
&=\tr \{ \E \{Ax_0 (x_0)^* A^*\} \} -2 \tr \{  B \E \{ x_0 (x_0)^* A^*\} \} \nonumber\\
& + \text{Tr} \{ B \E\{ x_0 (x_0)^* \} B^* \}.
\end{align}
Furthermore, for a linear system, $\Sigma_{\infty}$ satisfies the Lyapunov equation $A \Sigma_{\infty}  + \Sigma_{\infty} A^* + I = 0$.
For $A$ symmetric, this implies $\Sigma_{\infty} = -(1/2) A^{-1}$. 
Substituting this expression in \eqref{eq:norm2_err} and 
\eqref{eq:err_linear_est} finishes the proof.
\end{IEEEproof}

\subsection{Proof of Theorem \ref{th:linear_lbound_sparse}}

We prove the theorem by showing that the same complexity bound holds in the case when we are trying to estimate the signed support of $A$ for an $A$ that is uniformly randomly chosen with a distribution supported on ${\cA}^{(S)}$ and we simultaneously require that the average probability of error is smaller than $1/2$.
This guarantees that unless the bound holds, there will exist $A \in 
{\cA}^{(S)}$ for which the probability of error is biger than $1/2$.
The complexity bound for random matrices $A$ is proved using
Lemma \ref{th:mut_inf_bound_linear}. 

In order to generate $A$ at random we proceed as follows.
Let $G$ be the a random matrix constructed from 
the adjacency matrix of a uniformly random $k$-regular graph.
Generate $\tilde{A}$ 
by flipping the sign of each non-zero entry in  $G$ 
with probability $1/2$ independently. We  define 
$A$ to be the random matrix $A = -(\gamma + 2a_{\min}\sqrt{k-1})I + a_{\min} \tilde{A}$ where  $\gamma=\gamma(\tA)>0$ is the smallest value such that the 
maximum eigenvalue of $A$ is smaller than $-\rho$.
This guarantees that all these $A$ satisfy the four properties
of the class $\cA^{(S)}$.

The following lemma encapsulates the necessary random matrix calculations.
\begin{lemma} \label{th:random_matrix_calc_sparse}
Let $A$ be a random matrix defined as above and 
\begin{align}
Q(a_{\min},k,\rho)\equiv \lim_{p \rightarrow \infty} \frac{1}{p} \{ \tr \{ \E(-A) \}  - \tr \{ (\E(-A^{-1}))^{-1} \} \}.
\end{align}
Then, there exists a constant  $C'$ only dependent on $k$ such that
\begin{align}
&Q(a_{\min},k,\rho)  \leq \min \{\frac{C' k a^2_{\min}}{\rho} ,\frac{k a_{\min}}{\sqrt{k-1}}\}.
\end{align}
\end{lemma}
\begin{IEEEproof}
First notice that 
\begin{align} \label{eq:linear_bound_first_term}
&\lim_{p \rightarrow \infty } \frac{1}{p} \E \text{Tr} \{-A\} = \lim_{p \rightarrow \infty} \E(\gamma) + 2a_{\min}\sqrt{k-1} \\
&= \rho +  2a_{\min}\sqrt{k-1}
\end{align}
since by Kesten-McKay law \cite{friedman}, for large $p$, the spectrum of $\tilde{A}$ has support in $(- \epsilon -2a_{\min}\sqrt{k-1},2a_{\min}\sqrt{k-1} + \epsilon)$ with high probability. Notice that unless we randomize each entry of $\tilde{A}$ with $\{-1,+1\}$ values, every $\tilde{A}$ will have $k$ as its largest eigenvalue and the above limit will not hold.

For the second term we will compute a lower bound. For that purpose let $\lambda_i > 0$ be the $i^{th}$ eigenvalue of the matrix $\E(-A^{-1})$. We can write,
\begin{align} \label{eq:jensen_trick}
&\frac{1}{p} \text{Tr} \{ (\E(-A^{-1}))^{-1} \} = \frac{1}{p} \sum^p_{i = 1} \frac{1}{\lambda_i}\\
& \geq \frac{1}{\frac{1}{p} \sum^p_{i = 1} \lambda_i} =\frac{1}{ \E  \{ \frac{1}{p} \text{Tr} \{ (-A)^{-1}  \}\} }
\end{align}
where we applied Jensen's inequality in the last step. By Kesten-McKay law we now have that,
\begin{align}\label{eq:linear_bound_second_term}
& \lim_{p \rightarrow \infty} \E  \{ \frac{1}{p} \text{Tr} \{ (-A)^{-1}  \}\} = \E \{  \lim_{p \rightarrow \infty}    \frac{1}{p} \text{Tr} \{ (-A)^{-1}  \}   \} \\\
&= \frac{1}{a_{\min}} G(k,\rho/a_{\min}+2\sqrt{k-1})
\end{align}
where
\begin{align}
G(k,z) = \int \frac{-1}{\nu - z} \de \mu(\nu)\, 
\end{align}
and 
\begin{align} \label{eq:mckay_law}
\de \mu(\nu)  = \frac{k}{2 \pi} \frac{\sqrt{4(k-1) - \nu^2 }}{k^2 - \nu^2} \de \nu
\end{align}
for $\nu \in [-2\sqrt{k-1},-2\sqrt{k-1}]$ and zero otherwise. Expression \eqref{eq:mckay_law} defines the Kesten-McKay distribution. 
Computing the above integral we obtain
\begin{align}
G(k,z)  = -\frac{(k-2) z-k \sqrt{-4 k+z^2+4}}{2 \left(z^2-k^2\right)}\, 
\end{align}
whence
\begin{align}
 &\lim_{\rho \rightarrow 0} Q(a_{\min},k,\rho) = \frac{a_{\min} k}{\sqrt{k-1}}, \\
 &\lim_{\rho \rightarrow \infty}  \rho \, Q(a_{\min},k,\rho)  = k (a_{\min})^2. 
\end{align}
Since $Q(a_{\min},k,\rho)/a_{\min}$ is a function of $k$ and $\rho/a_{\min}$ that is strictly decreasing with $\rho/a_{\min}$, the claimed bound follows.
\end{IEEEproof}

\begin{IEEEproof}[Proof (Theorem \ref{th:linear_lbound_sparse})]
Starting from the bound of Lemma \ref{th:mut_inf_bound_linear}, we divide both terms in the numerator and the denominator by $p$. The term $H(M(A))/p$ can be lower bounded by $p^{-1} \log \left( \binom {p}{k} 2^k \right)^p \geq k \log(2 p/k)$ and Lemma \ref{th:random_matrix_calc_sparse} gives an upper bound on the denominator when $p \rightarrow \infty$. We now prove that $\lim_{p \rightarrow \infty} I(x_0;A)/p \leq 1$. This finishes the proof of Theorem \ref{th:linear_lbound_sparse} since after multiplying by a small enough constant (only dependent on $k$) the bound obtained by replacing the numerator and denominator with these limits will be valid for all $p$ large enough.

We start by writing,
\begin{align}
&I(x_0;A) = h(x_0) - h(x_0|A)  \label{eq:mut_inf_gaus_bound}\\
&\leq \frac{1}{2} \log (2 \pi e)^p |\E (\Sigma_{\infty})| - \E \frac{1}{2} \log (2 \pi e)^p | \Sigma_{\infty}| \label{eq:mut_inf_gaus_bound_1},
\end{align}
where $\Sigma_{\infty} = -(1/2) A^{-1}$ is the covariance matrix of the stationary process $x_t$ and $|.|$ denotes the determinant of a matrix. Then we write,
\begin{align}
I(x_0;A) &\leq \frac{1}{2} \log |\E (-(\beta A)^{-1})|  + \frac{1}{2} \E \log (|-\beta A|) \label{eq:mutual_info_intermediate}\\
&\leq \frac{1}{2} \text{Tr}\E (-I -(\beta A)^{-1})  + \frac{1}{2} \E \text{Tr}\{ -I-\beta A\}
\end{align}
where $\beta > 0$ is an arbitrary rescaling factor and the last inequality follows from $\log(I + M) \leq \text{Tr}(M)$. From this and equations \eqref{eq:linear_bound_first_term} and \eqref{eq:linear_bound_second_term} it follows that,
\begin{align}
\lim_{p \rightarrow \infty} \frac{1}{p} I(x_0;A) \leq -1 + (1/2) (\beta' z + \beta'^{-1} G(k,z))
\end{align}
where $z = \rho/ a_{\min} + 2 \sqrt{k-1}$ and $\beta' = \beta a_{\min} $. Optimizing over $\beta'$ and then over $z$ gives,
\begin{align}
\beta' z + \beta'^{-1} G(k,z) \leq 2 \sqrt{z G(k,z)} \leq \sqrt{8} \sqrt{\frac{k-1}{k-2}} \leq 4,
\end{align}
which implies $\lim_{p \rightarrow \infty} I(x_0;A)/p \leq 1$.
\end{IEEEproof}
%
%
\subsection{Proof of Theorem \ref{th:linear_lbound_dense}: Outline}
The proof of this theorem follows closely the proof of Theorem \ref{th:linear_lbound_sparse}. We will prove that same bound (\ref{eq:lbound_dense}) holds for an $A$ chosen at random with a distribution supported on $\cA^{(D)}$, whence the
claim follows. In order to lower bound the error probability for
random matrices, we make use of Lemma \ref{th:mut_inf_bound_linear}.

We construct the random matrix $A$ as follows.
Let $\tilde{A}$ be a random symmetric matrix with $\{A_{ij}\}_{i\le j}$
i.i.d. random variables where $\prob(A_{ij}=a_{\min}) = \prob(A_{ij}=-a_{\min})
=1/4$, and $\prob(A_{ij}=0) = 1/2$. Notice that the second moment of each entry is $\E(A_{ij}^2) = a^2_{\min}/2\equiv \alpha$. We then 
define $A = -(\gamma + 2 \sqrt{\alpha})I + \tilde{A}/\sqrt{p}$ where 
$\gamma=\gamma(\tA)$ is the smallest value that guarantees that $\lambda_{\min}(-A) \ge \rho$.

%
%
\subsection{Proof of Theorem \ref{th:non_linear_lbound}}

The proof consists in evaluating the lower bound in Corollary 
\ref{th:main_simpler_lbound}. We again prove the theorem by showing for a random class of functions contained in $\cA^{(N)}$.

We consider a the set of functions such that for each possible support of a $p$ by $p$ matrix with at most $k$ non-zero entries per row. Assume  there is one and only one function in the family with $JF$ having that support for all $x$.

Now notice that $\E_{x_0} \Var_{x_0|A} F(x_0;A) \leq \E (||F(x_0;A)||^2)$. Secondly notice that, if $x$ and $x'$ only differ on the $j^{th}$ component and $(JF)_{ij} \neq 0$ then $|F_i(x;A)| \leq |F_i(x';A)| + D||x'-x||$. Since $JF$ has at most $k$ non-zero entries per row, we get that for any $x$ and $x'$, $|F_i(x;A)| \leq |F_i(x';A)| + kD ||x'-x||$. If $x = x_0$ and $x' = \E_{x_0|A}(x_0|A)$ 
then squaring the previous expression and taking expectations gives us $\E_{x_0|A}(F_i(x;A)^2|A) \leq 2 F_i(x';A)^2 + 2k^2 D^2 B$. From this we get that $\E (||F(x_0;A)||^2)/p \leq C +  2k^2 D^2 B $ where $C$ is a constant independent of $A$.
For this sub family of functions we have $H(M(A)) \geq pk \log(p/k)$. By \eqref{eq:mut_inf_gaus_bound} and \eqref{eq:mut_inf_gaus_bound_1} we know that $I(x_0;A) \leq (1/2) \log((2 \pi e)^p |\E \Sigma_{\infty}| ) - (1/2) \E \log((2 \pi e)^p | \Sigma_{\infty}| )$. The first term, which is the entropy of a $p$-dimensional Gaussian with covariance matrix $\E \Sigma_{\infty}$, can be upper bounded by the sum of the entropy of its individual components, which have variance upper bounded by $B$. Finally, since $\Lambda_{\min}(\Sigma_{\infty}) \geq L$, we have $\log | \Sigma_{\infty}| \geq p \log L$ and therefore $I(x_0;A) \leq p/2 \log B/L$, which completes the proof.
\endproof

\subsubsection*{\bf Acknowledgments}

This work was partially supported by the NSF CAREER award CCF-0743978,
the NSF grant DMS-0806211, the AFOSR grant FA9550-10-1-0360 and by a Portuguese Doctoral FCT fellowship.

\bibliographystyle{IEEEtran}
\bibliography{inference}

\newpage

\appendix

\subsection{Proof of Theorem \ref{th:linear_lbound_dense}}

The following Lemma contains a matrix theory calculation that will be later used in this proof when applying Lemma \ref{th:mut_inf_bound_linear}. Recall that we defined $\alpha = a^2_{\min}/2$.

\begin{lemma} \label{th:random_matrix_calc_dense}
Let $A$ be a random matrix defined as above and 
\begin{align}
Q(a_{\min},\rho)\equiv \lim_{p \rightarrow \infty} \frac{1}{p} \{ \tr \{ \E(-A) \}  - \tr \{ (\E(-A^{-1}))^{-1} \} \}.
\end{align}
Then, there exists a constant  $C'$ such that
\begin{align}
&Q(a_{\min},\rho)  \leq \min \{\frac{C' a^2_{\min} }{2\rho} ,\frac{a_{\min}}{\sqrt{2}} \}.
\end{align}
\end{lemma}

\begin{IEEEproof}
Using Wigner's Semicircle law for random symmetric matrices \cite{Zeitouni} and the bound described in \eqref{eq:jensen_trick} it follows that,
\begin{align}
 &\lim_{p \rightarrow \infty} \frac{1}{p} \{ \text{Tr} \{ \E(-A) \}  = \rho + 2\sqrt{\alpha},\\
 &C(\alpha,\rho) \equiv \lim_{p \rightarrow \infty}  \E \{ \frac{1}{p} \text{Tr} \{ (-A)^{-1}\} \}  \label{eq:second_term_wigner}\\
&=\frac{-\sqrt{\rho  \left(4 \sqrt{\alpha }+\rho \right)}+2 \sqrt{\alpha }+\rho }{2 \alpha}.
\end{align}
Since $C(\alpha,\rho) = \alpha^{-1/2} C(1,\rho/\sqrt{\alpha})$ we can write 
$\rho +  2\sqrt{\alpha} - ( C(\alpha,\rho) )^{-1} = \sqrt{\alpha} G(\rho/\sqrt{\alpha})$ where $G(x)$ is a strictly decreasing function. Since $\lim_{\rho \rightarrow 0} = \sqrt{\alpha} G(\rho/\sqrt{\alpha}) = \sqrt{\alpha}$ and $\lim_{\rho \rightarrow \infty}  \rho \sqrt{\alpha} G(\rho/\sqrt{\alpha}) = \alpha$ it follows that there is a constant $C'$ independent of $\alpha$ or $\rho$ such that $\sqrt{\alpha} G(\rho/\alpha) \leq \sqrt{\alpha}\min \{1,C' \sqrt{\alpha} / \rho \}$. The result now follows by replacing $\alpha = a^2_{\min}/2$.
\end{IEEEproof}

\begin{IEEEproof}[Proof (Theorem \ref{th:linear_lbound_dense})]
Like in the proof of Theorem \ref{th:linear_lbound_sparse} we start by dividing both numerator and denominator of \eqref{eq:bound_linear} in Lemma 
\ref{th:mut_inf_bound_linear} by $p$. By multiplying the resulting expression by an appropriately small constant we can replace the denominator and $\lim_{p \rightarrow \infty} I(x_0;A)/p$ by their limits when $p \rightarrow \infty$ and get an expression that is still valid for all $p$ large enough. Since $H(M(A))/p = \frac{(1+p))}{4}  \log 4$, and since by Lemma \ref{th:random_matrix_calc_dense} we already know the limiting expression of the denominator, all we have to do is find $\lim_{p \rightarrow \infty} I(x_0;A)/p$. By an analysis very similar to that in the proof of Theorem \ref{th:linear_lbound_sparse} one can show that
\begin{align}
\lim_{p \rightarrow \infty} \frac{1}{p} I(x_0;A)  \leq -1 + \sqrt{(z+2) C(1,z)} \leq 1.
\end{align}
where $C(\alpha,\rho)$ was defined in \eqref{eq:second_term_wigner}, which finishes the proof.
\end{IEEEproof}

\end{document}